\documentclass{CUP-JNL-EDS}

\usepackage{latexsym}
\usepackage{graphicx}
\usepackage{multicol,multirow}
\usepackage{amsmath,amssymb,amsfonts}
\usepackage{mathrsfs}
\usepackage{amsthm}
\usepackage{rotating}
\usepackage{appendix}
\usepackage[authoryear]{natbib}
\usepackage{ifpdf}
\usepackage[T1]{fontenc}
\usepackage{times}
\usepackage{newtxmath}
\usepackage{textcomp}%
\usepackage{xcolor}%
\usepackage{hyperref}
\usepackage{lipsum}
\usepackage{algorithm}
\usepackage{algpseudocode}

\articletype{RESEARCH ARTICLE}
\jname{Environmental Data Science}

\begin{document}

\begin{Frontmatter}

\title[Article Title]{The impact of feature engineering and an optimisation framework for ocean colour machine learning}

\author*[1]{Edson Silva}\orcid{0000-0002-1097-9801}\email{edson.silva@nersc.no}
\author[1]{Julien Brajard}
\author[2]{Simon Cappe}
\author[2]{Lasse H. Pettersson}
\author[1]{François Counillon}

\address[1]{\orgname{Nansen Environmental and Remote Sensing Center, and Bjerknes Centre for Climate Research}, \orgaddress{\city{Bergen}, \state{Vestland}, \country{Norway}}}

\address*[2]{\orgname{Nansen Environmental and Remote Sensing Center}, \orgaddress{\city{Bergen}, \state{Vestland}, \country{Norway}}}


\authormark{Silva et al.}

\keywords{remote sensing, machine learning, ocean colour, feature engineering}

\abstract{Machine learning (ML)  is widely used for the development of ocean colour algorithms, but most studies focus on model parameter training and hyperparameter tuning.
The optimisation of the data that feeds the models --- i.e., Feature Engineering (FE) --- is not fully explored.
We assess the impact of FE in ocean colour machine learning models and we propose an optimisation framework that includes seven sequenced levels of data transformation: i. band choice, ii. log scaling, iii. spectral shape normalisation, iv. index extraction, v. principal component analysis, vi. feature scaling, and vii. zero-to-one scaling.
We demonstrate the application for Multi-layer perceptron, Support Vector Machines, and eXtreme Gradient Boosting Trees on Sentinel-3 OLCI observations in the Norwegian coastal waters.
The models are trained to estimate Chlorophyll-a concentration [Chl-a] and Secchi disk depth ($Z_{\text{sd}}$).
Results show that accuracy is highly variable among FE  found in six studies using Sentinel-3 OLCI and the ones that we optimise.
The $R$ range from 0.01 to 0.55 for [Chl-a] and from 0.15 to 0.68 for $Z_{\text{sd}}$, where the optimised FE shows the top results.
The ML models with optimised FE could also improve by two times the $R$ and reduce up to 63\% of the mean absolute error when compared to CHL\_OC4ME and CHL\_NN standard algorithms.
Nevertheless, no common optimised FE is found for all target variables and ML models, suggesting that FE optimisation is necessary for each application.
Therefore, our proposed framework can be key for improving the accuracy of water quality monitoring in coastal waters.
}

\policy{Machine learning (ML) is commonly used for satellite monitoring of water quality in coastal areas.
While most studies focus on improving ML parameters and hyperparameters, here we focus on optimising the data that the ML is trained on (i.e., feature engineering --- FE).
Accuracy varies widely across the FE found in the Sentinel-3 OLCI literature.
Hence, we propose an optimisation framework that employs a search algorithm in the FE space to find the best option for feeding the machine learning model.
These options include which bands to use, scaling transformations, and index extraction.
The optimised FE options depend on the machine learning model and target variable.
Nevertheless, applying this framework to each application yields more accurate retrieval models for satellite monitoring of water quality.}
\end{Frontmatter}

\section[Introduction]{Introduction}
Satellite ocean colour is challenging in coastal waters where the optically active constituents are not covarying and chlorophyll-a concentration [Chl-a] estimation by blue and green band ratios regularly fails \citep{pettersson_monitoring_2013}.
Consequently, other water quality parameters, such as Secchi depth ($Z_{\text{sd}}$), cannot be derived by the same [Chl-a] estimations \citep{morel_optical_2007}.
In these often-called Case-2 waters \citep{morel_analysis_1977}, a wide range of methodologies and approaches are employed for developing regionally tuned algorithms \citep{ioccg_remote_2000,ioccg_earth_2018}.
Recently, the use of machine learning (ML) for estimating bio-geochemical and water quality parameters has become popular in the remote sensing monitoring of coastal and inland waters \citep{brando_phytoplankton_2021, joshi_monitoring_2024, lapucci_use_2023, pahlevan_seamless_2020, shen_sentinel-3_2020, zoffoli_ciao_2025}.
The expansion of in-situ databases matching satellite observations fosters the increase of ML as it can easily exploit large databases to calibrate retrieval models.
Improving the models’ accuracy involves a training procedure to optimise the model parameters, but depends in large part on tuning hyperparameters, such as the type of ML architecture, the size of the model, among others.
The transformation of the data that feeds the models – i.e., Feature Engineering (FE) – also strongly affects ML models' accuracy \citep{gada_automated_2021, mumuni_automated_2025, verdonck_special_2024}, but its impact on ocean colour ML models is little understood and probably remains ad-hoc.
Ocean colour ML models frequently employ FE, such as band selection, remote sensing reflectance ($R_{rs}$) scaling, and computation of indexes  \citep{brando_phytoplankton_2021, lapucci_use_2023, pahlevan_seamless_2020, zoffoli_ciao_2025}, which can span a large number of possible unique options when combined.
Hence, feature engineering in ocean colour might benefit more by using optimisation techniques than a manual searching approach.

FE consists of preparing and transforming the data into features that an ML algorithm can train on.
In a broad definition, FE includes feature selection, univariate transformations, multivariate changes, and domain-specific FE \citep{verdonck_special_2024}.
Despite modern machine learning architectures --- e.g., deep learning --- being able to extract features during the model optimisation, they are still improved by FE techniques  \citep{fazliani_enhancing_2025, mumuni_automated_2025, heaton_automated_2017, heaton_empirical_2016}.
Besides, deep learning typically benefits most when large volumes of data are available \citep{rather_breaking_2024, safonova_ten_2023}, which is often not feasible in ocean colour ML.
Calibration relies on  in-situ water quality data matching radiometric measurements (in-situ or satellite), which are scarce due to field campaigns costs or geographical constraints for the calibration of regional models.
For example, regional algorithms for Sentinel-3 OLCI have been calibrated with the number of samples spanning from 367 to 2,943 \citep{brando_phytoplankton_2021, joshi_monitoring_2024, lapucci_use_2023, pahlevan_seamless_2020, shen_sentinel-3_2020, zoffoli_ciao_2025}.
Consequently, these studies use traditional machine-learning algorithms that can better fit small datasets, such as Optimizable Gaussian Regression Models \citep{zoffoli_ciao_2025}, Multi-Layer Perceptron (MLP) \citep{brando_phytoplankton_2021}, Random Forests \citep{joshi_monitoring_2024, lapucci_use_2023, shen_sentinel-3_2020}, and Mixture Density Networks \citep{ pahlevan_seamless_2020}.
Traditional ML methods are more impacted by FE when compared to deep learning \citep{heaton_empirical_2016}, and, therefore, FE optimisation could greatly enhance ocean colour ML calibration.

Ocean colour FE often starts by choosing which bands to feed into the ML model.
Some studies use a few bands \citep{brando_phytoplankton_2021, shen_sentinel-3_2020, zoffoli_ciao_2025}, while others employ the whole optical spectrum \citep{joshi_monitoring_2024}, or only remove a few inadequate bands \citep{pahlevan_seamless_2020}.
The reason for selecting a few bands or using all of them is not always clear. 
We suppose that selecting a few bands may avoid over-fitting in data data-poor configuration, but can also be suboptimal.
Univariate transformations help improve normal distribution, which may increase convergence of the training, such as log scaling \citep{brando_phytoplankton_2021, pahlevan_seamless_2020, zoffoli_ciao_2025}, z-score scaling \citep{brando_phytoplankton_2021}, and inter-quartile scaling \citep{pahlevan_seamless_2020}.
These transformations accommodate samples skewed towards low $R_{rs}$ \citep{zoffoli_ciao_2025}, help against outliers, and enforce homogeneous scale across all bands \citep{pahlevan_seamless_2020}.
Multivariate transformations in ocean colour models, however, are lacking to the extent of our knowledge, but a commonly known method is Principal Component Analysis (PCA) \citep{verdonck_special_2024}.
Domain-specific FE in ocean colour has been little explored as well, but one example is combining maximum band ratio (MBR) with $R_{rs}$ to feed the ML models \citep{lapucci_use_2023}.
The band ratio may maximise the relationship with [Chl-a] and improve the models' accuracy \citep{lapucci_use_2023}.
Many more domain-specific FE could be considered, such as normalising the $R_{rs}$ \citep{soja-wozniak_novel_2017}, estimation of Slopes and Normalised Difference Chlorophyll Index (NDCI), as well as other band ratios (BR) \citep{cairo_hybrid_2020}.
All the aforementioned FE methods can stack in many possible combinations. \cite{pahlevan_seamless_2020} FE with Sentinel-3 OLCI, for example, starts with excluding the band at 400 nm, log scaling the $R_{rs}$, normalising using inter-quartile scaling, and scaling from 0 to 1.
Although different FE choices are probably explored during the ML model calibration, FE may remain little optimised considering the many possible combinations.
Nevertheless, the FE could be optimised by mirroring the optimisation of ML hyperparameters.

The hyperparameters of ML models span a variety of options and carefully tuning them impacts the accuracy of models.
For example, the hyperparameters of the Support Vector Machine regressor (SVM) include the kernel function, penalty factor, epsilon, and other kernel-dependent hyperparameters, such as gamma and degree \citep{cortes_support-vector_1995, drucker_support_1997}.
Finding the optimal hyperparameter configuration involves creating a search space and employing a search algorithm to find the optimal choice evaluated through cross-validation.
The search space dimension is given by the number of hyperparameters that span a variety of options.
Considering SVM, for example, the options can be few, such as the kernel among linear, radial basis, and sigmoid options, and many, such as any real number above 0 for the penalty factor.
Since the search space can extend to many options, Tree of Parzen Estimators (TPE) is a commonly used search algorithm that optimises hyperparameters.
This search algorithm is a sequential optimisation that uses Gaussian Mixture Models (GMM) to infer the following hyperparameters in a sequence of trials that maximise the likelihood of improving accuracy \citep{bergstra_algorithms_2011,bergstra_making_2013}.
Roughly speaking, the search can start with 50 random hyperparameter configurations.
Then, the best and worst hyperparameters are fitted in GMMs to infer the hyperparameters more likely to increase the accuracy in the next trial.
The optimisation stops when the number of trials reaches a given limit and the optimised hyperparameters are the trial that reached the highest accuracy among all.
TPE not only converges to optimal configurations faster but also creates ML models more accurate than randomly searching hyperparameters \citep{bergstra_making_2013}.

Considering that search algorithms are quite successful for optimising the hyperparameter choice by giving a hyperparameter space, our hypothesis is that we can create an FE space and employ the same search algorithm to optimise the FE choice.
Hence, our main objective is to assess whether optimising FE yields more accurate ML models in coastal waters for the retrieval of water quality.
Our study focuses on the Norwegian coastal and fjord waters, characterised by high colored dissolved organic matter (CDOM) concentration and coccolithophore blooms, and where open ocean (Case-1) optical algorithms perform poorly \citep{tessin_testing_2024}.
The target variables are the [Chl-a] and Secchi depth ($Z_{\text{sd}}$), and the predictors are the Sentinel-3 OLCI reflectance.
We focus on three ML models:  MLP, SVM, and eXtreme Gradient Boosting trees (XGBoost).
We calibrate the ML models by adapting FE found in previous studies \citep{brando_phytoplankton_2021, joshi_monitoring_2024, lapucci_use_2023, pahlevan_seamless_2020, shen_sentinel-3_2020, zoffoli_ciao_2025} to assess the impact of FE on the ML models' accuracy.
Then, we create an FE space to employ the TPE search algorithm and optimise the FE.
Finally, we choose the best model to demonstrate its application in satellite images and evaluate increased performance against Sentinel-3 OLCI standard algorithms.

\section[Material and Methods]{Material and Methods}
\subsection{Study region}
\begin{figure}[!]
\centering
\includegraphics[width=1\textwidth]{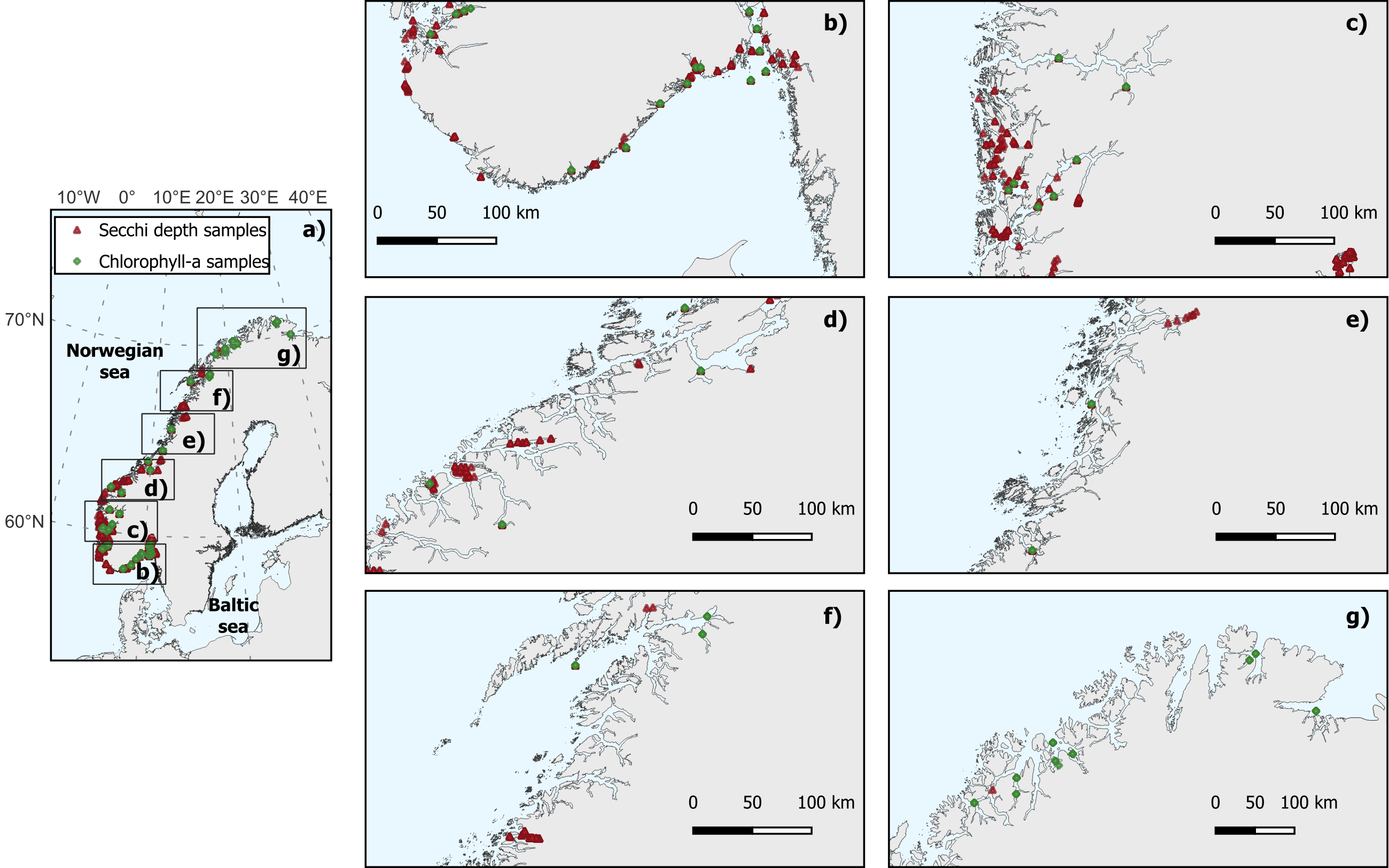}
\caption{The Norwegian coast and in-situ monitoring stations. Subplot a) shows the subdivision of regions that are zoomed in subplots b-g). Red triangles are the Secchi depth ($Z_{\text{sd}}$) stations and green circles are the Chlorophyll-a concentration [Chl-a] stations. Most stations have more than one repeated observation as they come from continuous monitoring programs (e.g., Økokyst).}
\label{fig:fig1}
\end{figure}

In Norwegian coastal waters, CDOM dominates the absorption at wavelengths shorter than 600 nm, followed consecutively by phytoplankton and non-algal particles (NAP) \citep{nima_absorption_2016}.
In southern Norway (Figure \ref{fig:fig1}), the inflow from the Baltic Sea and rivers increases the CDOM during the melting of snow in summer.
This rich CDOM waters flows through the Norwegian Coastal Current (NCC) and moves northwards along the coast until the Barents Sea Opening.
In the fjords of northern Norway, a rapid increase in freshwater inflow caused by storms may lead to a sudden increase in CDOM \citep{frigstad_influence_2020}.
Coccolithophore blooms are common in the fjords, NCC and open ocean areas and have a large influence on the back-scattering \citep{kondrik_tendencies_2017, tessin_testing_2024}.
Furthermore, the optical variability in $R_{rs}$, estimated by satellite sensors, is likely influenced by the complicated geometric light field in the region.
The sun zenith angle reaches more than 70 degrees towards the winter months, which reduces the total light backscattered to the orbital sensor and increases the signal-to-noise ratio.
The mountains around the fjords can potentially create adjacency effects in some areas in the estimated $R_{rs}$.
The low $R_{rs}$ and the ambient light field make the standard Sentinel-3 OLCI bio-optical algorithms of [Chl-a] perform poorly in the region \citep{tessin_testing_2024} and pose a challenging task for the calibration of regional algorithms.

\subsection{Sentinel-3 Ocean and Land Colour Instrument (OLCI) observations}
Sentinel-3 OLCI (satellite A) images matching the in-situ samples from national monitoring programs on the same day are accessed from 2016 to 2023.
We first access the Sentinel-3 OLCI L1B product with the bands in radiance in 21 bands centred at 400 nm to 1020 nm at 300 m spatial resolution. We then use the Polymer atmospheric correction \citep{steinmetz_atmospheric_2011} to retrieve the surface reflectance, as they are shown to perform slightly better than other atmospheric corrections in Norwegian fjord waters \citep{tessin_testing_2024}.
Polymer employs a spectral matching method relying on a polynomial atmospheric model that encompasses the atmospheric scattering and sun-glint and a bio-optical ocean water reflectance model \citep{steinmetz_atmospheric_2011}.
For Sentinel-3 OLCI, the ocean water reflectance model accounts for coastal Case-2 waters and works continuously with open ocean waters \citep{steinmetz_sentinel-2_2018}.
We use the Polymer version v4.17beta2 and set it to use the NASA auxiliary data for the atmospheric information and correction.
We follow the suggested bitmask quality flagging for valid pixels (bitmask \& 1023 == 0) and we remove too bright pixels where the Near-infrared (NIR) reflectance is above 0.1.

The Sentinel-3 OLCI reflectance is extracted from a 3x3 pixel average over each match-up.
Since some stations are close to the coastline, some 3x3 window retrieves a few pixels masked as land.
About 60\% of all match-ups have the whole 3x3 pixel area with valid reflectance estimations.
Removing the remaining 40\% match-ups would largely affect the quality of ML models and also create a geographical bias, e.g. in the narrow fjord areas, which are important to be represented in the model.
Hence, we decided to use all match-ups with a minimum of one pixel available within the 3x3 window.

We assess Sentinel-3 L2B Non-Time Critical (NT) [Chl-a] products estimated by the CHL\_OC4ME and CHL\_NN algorithms and use them as a benchmark for comparing the improvements reached by our FE optimisation framework.
Both CHL\_OC4ME and CHL\_NN are standard products of Sentinel-3 OLCI level 2 available on the EUMETSAT catalogue and thus a suitable benchmark.
The CHL\_OC4ME algorithm uses the Baseline Atmospheric Correction (BAC) and maximum band ratio algorithm that is designed for estimating [Chl-a] in Case-1 waters \citep{eumetsat_sentinel-3_2021}.
The CHL\_NN is a C2RCC product that estimates [Chl-a] using neural networks with input the reflectance on the top of the atmosphere \citep{eumetsat_sentinel-3_2021, glukhovets_evaluation_2020, schiller_neural_1999}.
Furthermore, we create a $Z_{\text{sd}}$ benchmark by applying the following equation \citep{morel_optical_2007}:

\begin{equation}\tag{1}
Z_{\_sd} = 8.5 - 12.6\,\log_{10}\!\big([{\rm Chl\text{-}a}]\big)
+ 7.36\,\log_{10}\!\big([{\rm Chl\text{-}a}]\big)^{2}
- 1.43\,\log_{10}\!\big([{\rm Chl\text{-}a}]\big)^{3}
\label{eq:eq1}
\end{equation}
where $Z_{\text{sd}}$ (m) is estimated by using [Chl-a] from CHL\_OC4ME and CHL\_NN.

\subsection{Characterisation of optical water variability}
The Sentinel-3 OLCI observations matching the \textit{in-situ} data are clustered into Optical Water Types (OWTs). 
We characterise the water masses by OWTs separately for [Chl-a] and $Z_{\text{sd}}$ match-ups. 
Only for the OWT characterisation, we divide reflectance by $\pi$ to retrieve the remote sensing reflectance ($R_{rs}$) for comparison with previous studies \citep{tessin_testing_2024}. 
The K-means algorithm with the K-means++ initialisation \citep{arthur_k-means_2007} is used to cluster the 16 Sentinel-3 OLCI ocean colour bands from 400~nm to 1020~nm. 
We set the number of OWTs---i.e., clusters---to 9. 
Note that we use OWTs for describing water masses, and they are not considered in the calibration or optimisation of models.

\subsection{In-situ Chlorophyll-a and Secchi disk depth observations}
In-situ [Chl-a] and Secchi disk depth ($Z_{\text{sd}}$) are retrieved from several sources available on a public repository.
The [Chl-a] measurements are from Økokyst, a Norwegian ecosystem monitoring program for assessing the water quality of the Norwegian coastal waters, commissioned by the Norwegian Environment Agency.
The program runs in cycles every 4 years, with annual reports on the ecosystem status split into several sub-regions along the Norwegian coast.
The sampling is made at several permanent locations along the fjords that are repeated every month (Figure \ref{fig:fig1}).
The [Chl-a] is made available at the Vannmiljø portal (https://vannmiljo.miljodirektoratet.no/), a web portal from the Norwegian Environment Agency that comprises data retrieved from several monitoring projects.
The samples cover different depths, and only the stations with surface samples at the upper 10 m are utilised and averaged when more than one sample is present.
We retrieved 625 match-up samples where 506 follow the NS-ISO 5667-9A protocol method, 12 are measured on ferry boxes (automated measurements onboard commercial vessels), and 107 without measurement method description.
Since the samples are used for the official assessment of the coastal ecosystem status in Norway, and follow good quality control as defined in Classification of environmental state in water \citep{direktoratsguppen_vanndirektivet_2018_klassifisering_2018}, we have also used the 107 samples without a method description to maximise the number of samples.

Measurements of $Z_{\text{sd}}$ are retrieved from several sources and commissioned monitoring programs that make their data available through the Vannmiljø web portal.
Since $Z_{\text{sd}}$ is little affected by different methodologies, we use all $Z_{\text{sd}}$ measurements available in the Vannmiljø web portal (Figure \ref{fig:fig1}) to maximise the number of samples. Hence, 1378 $Z_{\text{sd}}$ measurements are available for training and testing the models.

\subsection{Feature engineering}
\begin{figure}[!]
\centering
\includegraphics[width=1\textwidth]{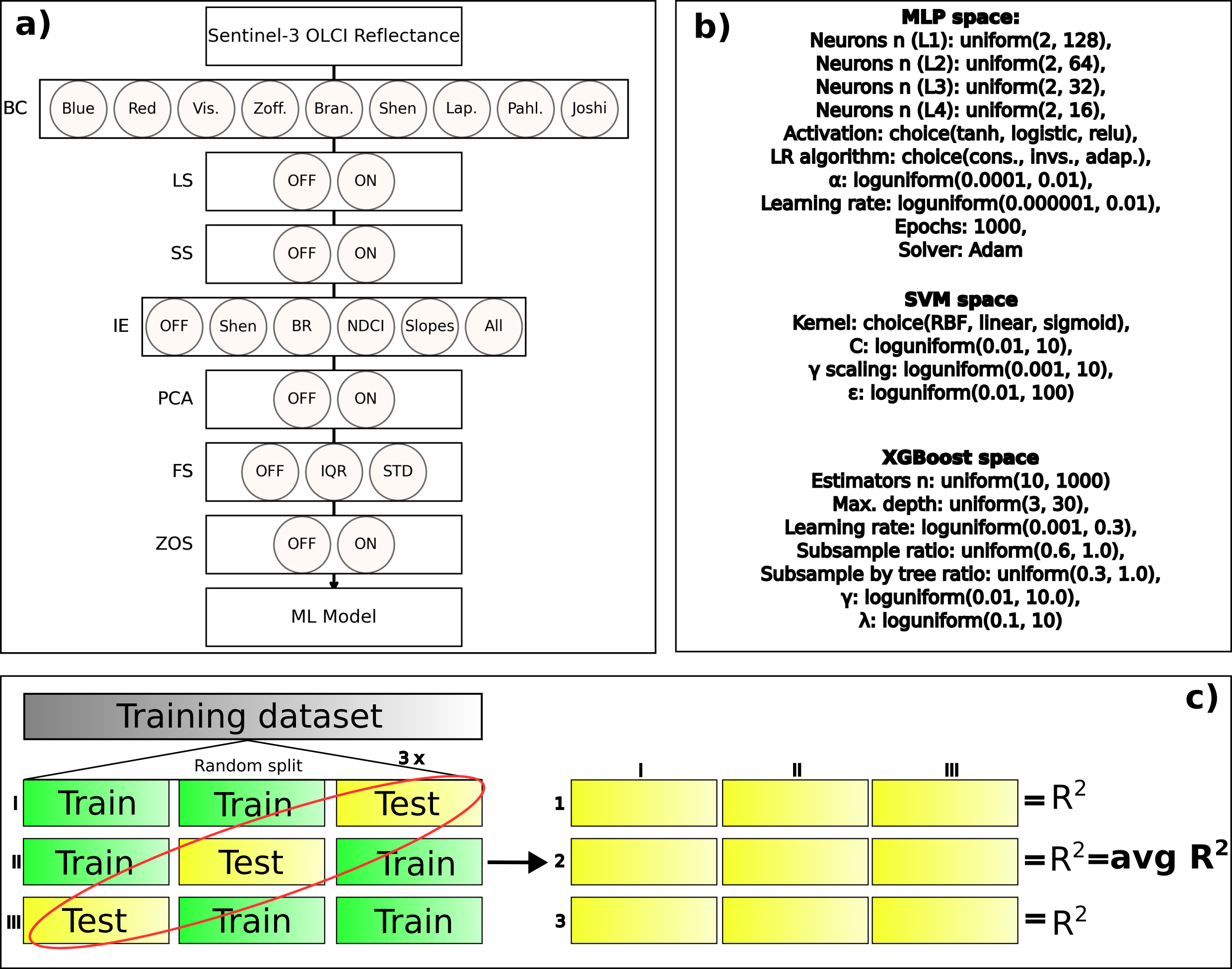}
\caption{Scheme of feature engineering (FE) and Machine Learning (ML) hyperparameters optimisation. Subplot a) exhibits the FE space where the Sentinel-3 OLCI reflectance feeds and is transformed across seven levels until reaching the ML model. Subplot b) shows the hyperparameter space for MLP, SVM, and XGBoost. Subplot c) shows the cross-validation method for estimating the $R^2$ of a given FE and ML hyperparameters configuration}
\label{fig:fig2}
\end{figure}

We build a Feature Engineering (FE) space consisting of seven successive levels of feature selection and data processing, where the data flows from the Sentinel-3 OLCI reflectance as input of the first level to the ML model input as the output of the seventh level (Figure \ref{fig:fig2}a).
The used Sentinel-3 OLCI reflectance bands are centred at wavelengths ($\lambda$) 400, 412.5, 442.5, 490, 510, 560, 620, 665, 673.75, 681.25, 708.75, 753.75, 778.75, 865, 885, and 1020 nm. 
The Bands Choice (BC) is the first level of FE and subsets the data depending on the choice among 8 different options, as described in Algorithm \ref{alg:alg1}.
The BC indicates the option (e.g., Red or Blue), X denotes the feature, and L1 refers to the output of level 1.
Since this is the first level, X is the Sentinel-3 OLCI reflectance.
The options Blue, Red, and Visible are created to select different or all bands in the visible spectrum, while the remaining options are adapted from recent studies following the name of the first authors \citep{brando_phytoplankton_2021, joshi_monitoring_2024, lapucci_use_2023, pahlevan_seamless_2020, shen_sentinel-3_2020, zoffoli_ciao_2025}.
Note that some studies use other input information in the ML models, such as day of the year \citep{zoffoli_ciao_2025} and bathymetry \citep{lapucci_use_2023}, but we have not included them as we want to limit the experiment to the optical calibration.
\cite{brando_phytoplankton_2021} employs an ensemble of ML models and we took only the option that showed the best accuracy. 

\begin{algorithm}[!]
\caption{Selection of $X_{L1}$ bands based on the BC option}
\begin{algorithmic}[1]
\Require BC option, Sentinel-3 OLCI Reflectance
\Ensure Selected band set $X_{L1}$
\If{BC = Blue}
    \State $X_{L1} \gets \{X_{400}, X_{412.5}, X_{442.5},X_{490}, X_{510}\}$
\ElsIf{BC = Red}
    \State $X_{L1} \gets \{X_{560}, X_{620}, X_{665},X_{673.75}, X_{681.25}, X_{708.75}\}$
\ElsIf{BC = Visible}
    \State $X_{L1} \gets \{X_{400}, X_{412.5}, X_{442.5}, X_{490}, X_{510}, X_{560}, X_{620}, X_{665}, X_{673.75}, X_{681.25}\}$
\ElsIf{BC = Zoffoli}
    \State $X_{L1} \gets \{X_{442.5}, X_{490}, X_{510}, X_{560}\}$
\ElsIf{BC = Brando}
    \State $X_{L1} \gets \{X_{490}, X_{510}, X_{560}, X_{673.75}\}$
\ElsIf{BC = Shen}
    \State $X_{L1} \gets \{X_{412.5}, X_{442.5}, X_{490}, X_{510}, X_{560}, X_{665}\}$
\ElsIf{BC = Lappucci}
    \State $X_{L1} \gets \{X_{412.5}, X_{442.5}, X_{490}, X_{510}, X_{560},X_{620}, X_{665}, X_{673.75}, X_{681.25}\}$
\ElsIf{BC = Pahlevan}
    \State $X_{L1} \gets \{X_{412.5}, X_{442.5}, X_{490}, X_{510}, X_{560},X_{620}, X_{665}, X_{673.75}, X_{681.25}, X_{708.75},X_{753.75}, X_{778.75}\}$
\ElsIf{BC = Joshi}
    \State $X_{L1} \gets \{X_{400}, X_{412.5}, X_{442.5}, X_{490}, X_{510},X_{560}, X_{620}, X_{665}, X_{673.75}, X_{681.25}, X_{708.75}, X_{753.75}$
    \State \quad $ X_{778.75}, X_{865}, X_{885}, X_{1020}\}$
\EndIf
\end{algorithmic}
\label{alg:alg1}
\end{algorithm}

The second level is the log scaling (LS) that log-scales the Sentinel-3 OLCI reflectance as described in Algorithm \ref{alg:alg2}.
Note that the 0.03 constant is necessary as negative reflectance is present in a few observations. 

\begin{algorithm}[!]
\caption{Computation of $X_{L2,\lambda}$ based on the LS option}
\begin{algorithmic}[1]
\Require LS option, $X_{L1,\lambda}$
\Ensure Computed $X_{L2,\lambda}$
\If{LS = ON}
    \State $X_{L2,\lambda} \gets \log_{10}(X_{L1,\lambda} + 0.03)$
\ElsIf{LS = OFF}
    \State $X_{L2,\lambda} \gets X_{L1,\lambda}$
\EndIf
\end{algorithmic}
\label{alg:alg2}
\end{algorithm}

The third level is the Spectral Shape (SS) normalisation and is described in Algorithm \ref{alg:alg3}. The summing is over all bands on the same sample that depends on the subset bands at level BC. 

\begin{algorithm}[!]
\caption{Computation of $X_{L3,\lambda}$ based on the SS flag}
\begin{algorithmic}[1]
\Require SS option, $X_{L2,\lambda}$
\Ensure Computed $X_{L3,\lambda}$
\If{SS = ON}
    \State $X_{L3,\lambda} \gets X_{L2,\lambda} \;/\; \sum X_{L2}$
\ElsIf{SS = OFF}
    \State $X_{L3,\lambda} \gets X_{L2,\lambda}$
\EndIf
\end{algorithmic}
\label{alg:alg3}
\end{algorithm}

The fourth level is the Index Extraction (IE), which gives several options that include the band ratios (BR), Normalised Difference of Chlorophyll Index (NDCI), Slopes, and Maximum Band Ratio (MBR). IE is described in Algorithm \ref{alg:alg4}.
Note that f means feature and replace $\lambda$ at level 4 as the observations can include features without $\lambda$.

\begin{algorithm}[!]
\caption{Computation of $X_{L4,f}$ based on the IE option}
\begin{algorithmic}[1]
\Require IE option, $X_{L3,\lambda}$
\Ensure Computed $X_{L4,f}$
\If{IE = BR}
    \State $X_{L4,f} \gets X_{L3,\lambda} \parallel (X_{L3,442.5}/X_{L3,560}) \parallel (X_{L3,442.5}/X_{L3,510})$
    \State \quad $\parallel (X_{L3,442.5}/X_{L3,400}) \parallel (X_{L3,560}/X_{L3,665}) \parallel (X_{L3,673.75}/X_{L3,665}) \parallel (X_{L3,681.25}/X_{L3,665})$
\ElsIf{IE = NDCI}
    \State $X_{L4,f} \gets X_{L3,\lambda} \parallel \frac{X_{L3,673.75}-X_{L3,665}}{X_{L3,673.75}+X_{L3,665}} \parallel \frac{X_{L3,681.25}-X_{L3,665}}{X_{L3,681.25}+X_{L3,665}} \parallel \frac{X_{L3,560}-X_{L3,665}}{X_{L3,560}+X_{L3,665}}$
\ElsIf{IE = Slope}
    \State $X_{L4,f} \gets X_{L3,\lambda} \parallel \frac{X_{L3,673.75}-X_{L3,665}}{673.75-665} \parallel \frac{X_{L3,681.25}-X_{L3,665}}{681.25-665} \parallel \frac{X_{L3,560}-X_{L3,665}}{665-560}$
\ElsIf{IE = MBR}
    \State $X_{L4,f} \gets X_{L3,\lambda} \parallel (X_{L3,442.5}/X_{L3,510}) \parallel (X_{L3,490}/X_{L3,510}) \parallel (\max(X_{L3,442.5}, X_{L3,490}) / X_{L3,510})$
\ElsIf{IE = All}
    \State $X_{L4,f} \gets X_{L3,\lambda} \parallel \text{BR} \parallel \text{NDCI} \parallel \text{Slope} \parallel \text{MBR}$
\ElsIf{IE = OFF}
    \State $X_{L4,f} \gets X_{L3,\lambda}$
\EndIf
\end{algorithmic}
\label{alg:alg4}
\end{algorithm}

The fifth level is the PCA and is described in Algorithm \ref{alg:alg5}.
The PCA computes the eigenvectors and eigenvalues for the training dataset, while the test dataset is only projected onto the eigenvectors derived from the training data.
The resulting transformed test data retains the variance structure defined by the training eigenvectors.

\begin{algorithm}[!]
\caption{Computation of $X_{L5,f}$ based on the PCA option}
\begin{algorithmic}[1]
\Require PCA flag, $X_{L4,f}$
\Ensure Computed $X_{L5,f}$
\If{PCA = ON}
    \State $X_{L5,f} \gets \text{PCA}(X_{L4,f})$
\ElsIf{PCA = OFF}
    \State $X_{L5,f} \gets X_{L4,f}$
\EndIf
\end{algorithmic}
\label{alg:alg5}
\end{algorithm}

The feature scaling is the sixth level and can apply scaling by the standard deviation (STD) and inter-quartile range (IQR) as described in Algorithm \ref{alg:alg6}.

\begin{algorithm}[!]
\caption{Computation of $X_{L6,f}$ based on the FS option}
\begin{algorithmic}[1]
\Require FS flag, $X_{L5,f}$
\Ensure Computed $X_{L6,f}$
\If{FS = STD}
    \State $X_{L6,f} \gets \frac{X_{L5,f,s} - \text{MEAN}(X_{L5,f})}{\text{STD}(X_{L5,f})}$
\ElsIf{FS = IQR}
    \State $X_{L6,f} \gets \frac{X_{L5,f,s} - \text{MEDIAN}(X_{L5,f})}{\text{IQR}(X_{L5,f})}$
\ElsIf{FS = OFF}
    \State $X_{L6,f} \gets X_{L5,f}$
\EndIf
\end{algorithmic}
\label{alg:alg6}
\end{algorithm}

The zero-to-one scaling (ZOS) is the last level and scales the data between 0 and 1 for using the minimum and maximum range of the feature as described in Algorithm \ref{alg:alg7}. Here, this is the last output and what is fed to the ML models.

\begin{algorithm}[!]
\caption{Computation of $X_{L7,f}$ based on the ZOS option}
\begin{algorithmic}[1]
\Require ZOS flag, $X_{L6,f}$
\Ensure Computed $X_{L7,f}$
\If{ZOS = ON}
    \State $X_{L7,f} \gets \frac{X_{L6,f} - \min(X_{L6,f})}{\max(X_{L6,f}) - \min(X_{L6,f})}$
\ElsIf{ZOS = OFF}
    \State $X_{L7,f} \gets X_{L6,f}$
\EndIf
\end{algorithmic}
\label{alg:alg7}
\end{algorithm}

\subsection{Machine learning models}
We employ MLP, SVM, and XGBoost models for covering a variety of ML architectures, as FE can depend on the ML model.
MLP is based on neural networks, SVM is based on support vectors, and XGBoost is based on decision trees. 
MLP is a type of artificial neural network that consists of several blocks (layers) of interconnected neurons that are linearly combined before being passed through a nonlinear function (activation).
During training, the model learns the coefficients of the weighted sum performed in each layer by minimising a loss function that quantifies the error between predicted and true values.
The optimisation techniques are often based on the gradient descent algorithm, with gradients efficiently computed through back-propagation. When samples are processed during optimization, the amount by which the coefficients are updated depends on the learning rate.
One epoch is when all samples are passed as batches through the optimisation function.
Hence, the MLP hyperparameters needed to be optimised are the number of neurons in each layer, the activation function, the learning rate, and the regularisation strength (Figure \ref{fig:fig2}).
A detailed explanation of MLPs can be found in \cite{hinton_connectionist_1989}.

Support Vector Machines (SVM) regressor is the generalisation of the SVM. The kernel function configures the hyperplane to fit the samples, and popular choices are linear, radial basis function (RBF), and sigmoid.
For a given kernel function, the SVM finds the hyperplane that holds the maximum number of observations within a tube of width $\varepsilon$ to solve the optimisation problem.
Samples falling inside the $\varepsilon$-width tube contribute 0 to the optimisation.
The penalty factor $C$ is a regularisation parameter in the optimisation function that controls the trade-off between minimising the error and the complexity of the model.
Finally, the RBF kernel introduces a hyperparameter $\gamma$, which controls the influence of a single sample.
A smooth $\gamma$, in our experience, gives well-adjusted ML models and can be obtained as a function of the input variance: $\gamma_{\text{smooth}} = 1 / (n_{\text{features}} \times \text{VAR}(X_{L7}))$, where $n_{\text{features}}$ is the number of features and $\text{VAR}$ is the variance estimator.
Hence, we introduce a scaling factor $\gamma_{\text{scaling}}$ to estimate deviations from $\gamma_{\text{smooth}}$, given by $\gamma = \gamma_{\text{smooth}} \times \gamma_{\text{scaling}}$.
During hyperparameter tuning, we optimise $\gamma_{\text{scaling}}$ rather than the absolute value of $\gamma$.
A deeper explanation of SVMs and regression tasks can be found in \cite{cortes_support-vector_1995} and \cite{drucker_support_1997}.

XGBoost (for eXtreme Gradient Boosting) is a version of the gradient boosted trees algorithm.
It is an ensemble of classification and regression trees (CART) that can be applied for classification and regression tasks similar to Random Forest.
However, the key difference is in how the trees are trained. Gradient boosting refers to sequentially adding weak estimators (CART) that are trained to correct the errors of their predecessors until the maximum number of boosting rounds, or a desired level of accuracy is reached.
The term eXtreme comes from the strong regularisation that includes the number of leaves (gamma) in each CART and the terms on weights (lambda).
Hence, XGBoost hyperparameters include the number of estimators, gamma, lambda, and learning rate, which is multiplied by the output of each sequential tree.
Furthermore, other hyperparameters include the minimum and maximum depth of the trees, sub-sampling rate, and feature sampling rate.
More details are found in \cite{chen_xgboost_2016}.

\subsection{Optimisation of feature engineering and ML hyperparameters}
The FE and ML hyperparameters are combined in a single space – hereafter referred to as pipeline space – for optimising both concurrently.
The optimisation only uses the train dataset for [Chl-a] (2016-2019) and Z\textsubscript{sd} (2016-2020).
First, the training dataset is randomly divided into 3 splits, each yielding 3 folds (Figure \ref{fig:fig2}a).
One fold is held for validation, while the two remaining folds are for fitting the ML models.
The three validation folds are combined in a single validation dataset where the coefficient of determination (R\textsuperscript{2}) is estimated.
The R\textsuperscript{2} is computed for each of the random splits and a final averaged R\textsuperscript{2} is used for ranking the pipeline options.

The TPE optimisation starts with 20 trials using random pipeline configurations by using probability distributions of each configuration space.
The initial probability distributions are nominal (e.g., all FE options), discrete (e.g., number of neurons in MLP), and continuous (e.g., C in SVM).
For the categorical pipeline configurations, the space is the given optional categories with a uniform probability distribution in each option, such as Red and Blue for the BC FE.
For discrete and continuous configurations in the hyperparameter space, the space is defined by a minimum and a maximum in a uniform or log-uniform distribution.
In the TPE jargon, the categorical, uniform, and log uniform are described as choice, uniform, and log-uniform spaces (Figure \ref{fig:fig2}b).
Hence, the space for initially ranking the configurations is given and described for FE in Figure 2a and ML hyperparameters in Figure \ref{fig:fig2}b.

The initial 20 random configurations are ranked and split into the best group for the top ranking, where \textit{N} is the number of trials (\textit{N}=20 at the start), and the worst group with the remaining trials.
For each pipeline choice – FE and ML hyperparameter choices – TPE fits a Gaussian Mixture Model (GMM) for the best group with the prior optimised configuration  and another for the worst group.
Then, 24 candidates are drawn from  and the one with the maximum value is chosen as the configuration for the next trial.
In other words, the TPE algorithm selects the candidate by balancing exploitation (high probability under the good distribution) with exploration (low probability under the bad distribution).
These optimisations progress one by one until reaching 500 trials.
The best configuration among all those trials is chosen as the optimal configuration.

\subsection{Experiments design}
We draw several experiments for assessing the impact of FE choices and FE optimisation using TPE.
For the experiment named “Optimised FE”, we employ TPE for the whole pipeline space, which means that we optimise the FE and ML hyperparameters concurrently.
For the remaining experiments, the FE space is constrained by the configuration found in the literature and TPE optimises only the ML hyperparameters.
The FE experiments are described in Table \ref{tab:FE_config}.
For all experiments, the TPE optimisation is run 30 times to assess the accuracy uncertainty that resulted from the TPE.
Furthermore, the optimal FE frequency in the FE optimised experiment is estimated through the relative frequency of the FE choice in the 30 Optimised FE TPE runs.

\begin{table}[h]
\centering
\begin{tabular}{|l|l|}
\hline
\textbf{Experiment name} & \textbf{FE configuration} \\
\hline
Optimised FE & TPE is employed to the whole FE space. \\
\hline
Zoffoli et al. (2025) FE & BC = Zoff. and LS = ON. \\
\hline
Brando et al. (2021) FE & BC = Bran., LS = ON, and FS = STD. \\
\hline
Joshi et al. (2024) FE & BC = Joshi. \\
\hline
Pahlevan et al. (2020) FE & BC = Pahl., LS = ON, FS = IQR, and ZOS = ON. \\
\hline
Lapucci et al. (2023) FE & BC = Lapp. and IE = Lapp. \\
\hline
Shen et al. (2020) FE & BC = Shen. \\
\hline
\end{tabular}
\caption{Feature engineering (FE) configurations for the different experiments. Omitted options are set to OFF.}
\label{tab:FE_config}
\end{table}

\subsection{Testing of the models}
The test dataset for [Chl-a] ranges from 2021 to 2023, and Z\textsubscript{sd} ranges from 2022 to 2023 and comprises 38\% and 45\% of total match-ups, respectively.
The number of Z\textsubscript{sd} observations is largely biased to recent years, and hence we need to shift one extra year for training while reducing one for testing.
Setting the test dataset to the recent years helps to avoid any potential geographical and temporal bias \citep{mcgovern_why_2022} caused by autocorrelation of samples between the train and test datasets.
Furthermore, the temporal split considers whether the calibrated ML models still remain accurate in any potential recent changes in the optical variability.
For example, darkening of coastal waters has been observed in the Norwegian coast in the last decades \citep{aksnes_coastal_2009}.
The statistical evaluation employed includes the Pearson correlation ($R$), the Root Mean Squared Error (RMSE), and the Mean Absolute Percent Error (MAPE).
Note that $R$ is estimated using the targets log-scaled at base 10, while MAPE and RMSE are computed in the original scale of the target variable.
The Pearson correlation coefficient is defined as:
\begin{equation}
R = \frac{\sum_{i=1}^{n} (y_i - \bar{y})(\hat{y}_i - \bar{\hat{y}})}{\sqrt{\sum_{i=1}^{n} (y_i - \bar{y})^2} \sqrt{\sum_{i=1}^{n} (\hat{y}_i - \bar{\hat{y}})^2}},
\end{equation}
where $y_i$ and $\hat{y}_i$ denote the observed and predicted values, respectively, and $\bar{y}$ and $\bar{\hat{y}}$ are their means.

The Root Mean Squared Error is given by:
\begin{equation}
\text{RMSE} = \sqrt{\frac{1}{n} \sum_{i=1}^{n} (y_i - \hat{y}_i)^2},
\end{equation}

and the Mean Absolute Percent Error is expressed as:
\begin{equation}
\text{MAPE} = \frac{100}{n} \sum_{i=1}^{n} \left| \frac{y_i - \hat{y}_i}{y_i} \right|.
\end{equation}

\section{Results}
\subsection{Optical variability of Norwegian coastal waters and match-ups}
\begin{figure}[!]
\centering
\includegraphics[width=1\textwidth]{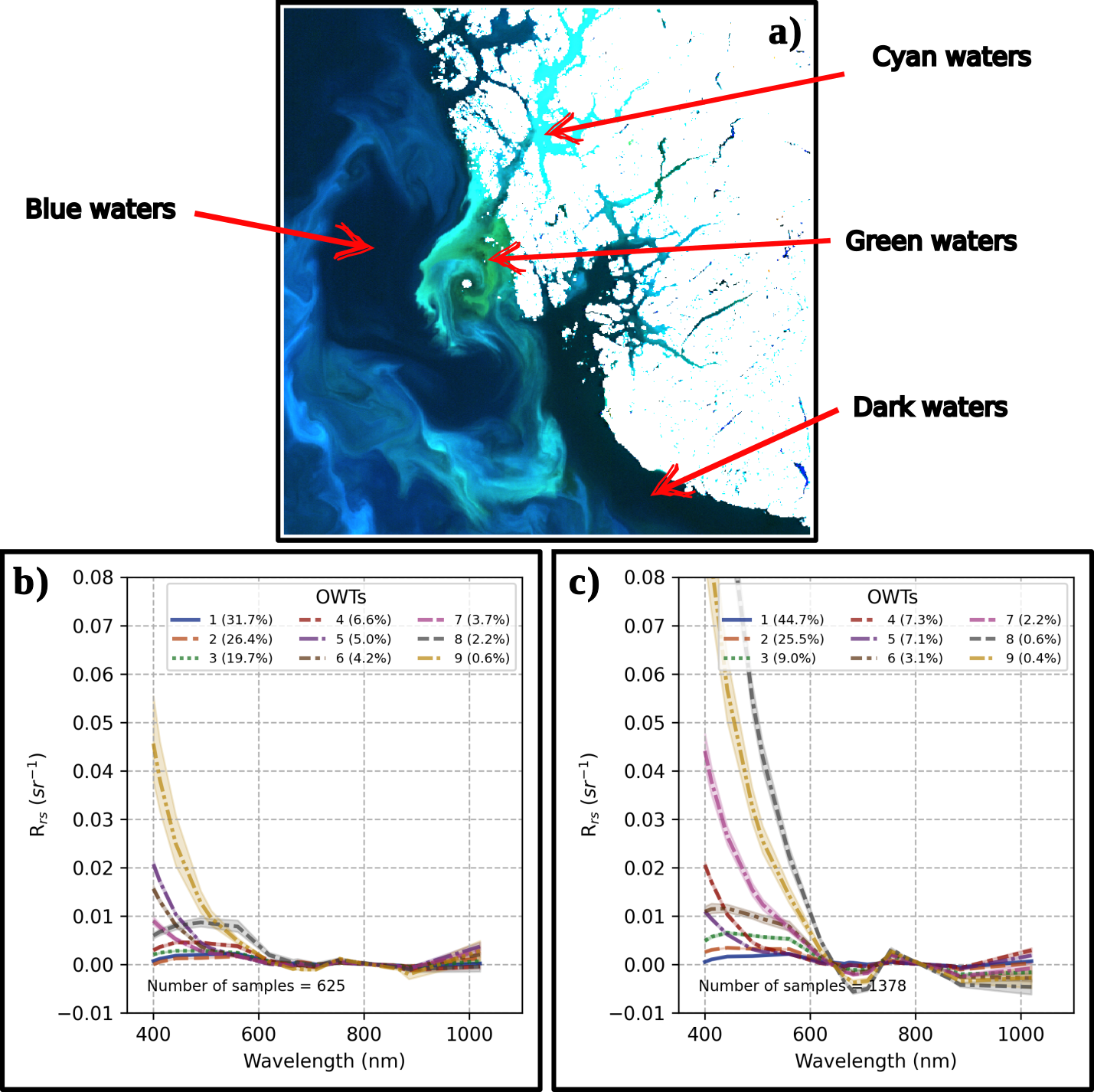}
\caption{Optical water type variability along the Norwegian coast and in-situ match-ups. Subplot a) is an example of a Sentinel-3 OCLI RGB-image (665 nm, 560 nm, and 443 nm) composition covering the southern  Norwegian coastal and fjord waters (2023-06-15) centred at 59.5°N and scaled with $R_{rs}$ between 0 and 0.01 sr$^{-1}$, resolving the colour contrast in the fjord-coastal waters. Subplot b) exhibits the estimated OWTs for the [Chl-a] match-ups (n = 625). Subplot c) shows the estimated OWTs for the $Z_{\text{sd}}$ match-ups (n = 1378)}
\label{fig:fig3}
\end{figure}

The Norwegian coastal waters exhibit a rich OWT variability that can be illustrated in a single OLCI scene (Figure \ref{fig:fig3}a).
Along the southern coast and in the fjords, dark waters with low R\textsubscript{rs} occur in the NCC.
Blue waters occur towards the open ocean and, in some inner fjords, high scattering waters in the green and blue range (cyan colour) appear.
Green waters are also present and occur along the coastline.
Dark waters are likely the CDOM-rich waters that are often carried by the NCC or more common through episodic high freshwater inputs in northern Norway \citep{frigstad_influence_2020}.
Blue waters are the oligotrophic open ocean waters (Case-1) with little influence from the terrestrial input of CDOM and sediments.
The high scattering and cyan waters are probably coccolithophore blooms that often occur in the fjords, by the coast and in the open ocean \citep{tessin_testing_2024}.
Green waters could be non-coccolithophore blooms that occur during the summer and are potentially harmful \citep{pettersson_monitoring_2013, john_spatial_2022}.

Considering the [Chl-a] match-ups (Figure \ref{fig:fig3}b), dark OWTs (1-3) with R\textsubscript{rs} average lower than 0.002 are 67.8\% of observations.
Cyan and green like OWTs (4 and 8) with relatively high R\textsubscript{rs} values are 8.8\% of the retrieved matches.
OWTs with higher R\textsubscript{rs} in blue (5, 6, 7, and 9), which indicate little influence of CDOM and blooms, are 13\% of the observations.
The OWTs observed in the Z\textsubscript{sd} match-ups exhibit a similar distribution but with the ones related to a blue shape exhibiting negative R\textsubscript{rs} around 700 nm (Figure \ref{fig:fig3}c).
Negative R\textsubscript{rs} also occur in a few samples in the blue region of the spectrum in dark waters in both [Chl-a] and Z\textsubscript{sd} match-ups (not shown).
Although they might present some constraints to physical models, from an ML point of view, they are not a problem as long as the reflectance variability corresponds to the optically active constituents.

\subsection{FE impact on ML models' accuracy}
\begin{figure}[!]
\centering
\includegraphics[width=1\textwidth]{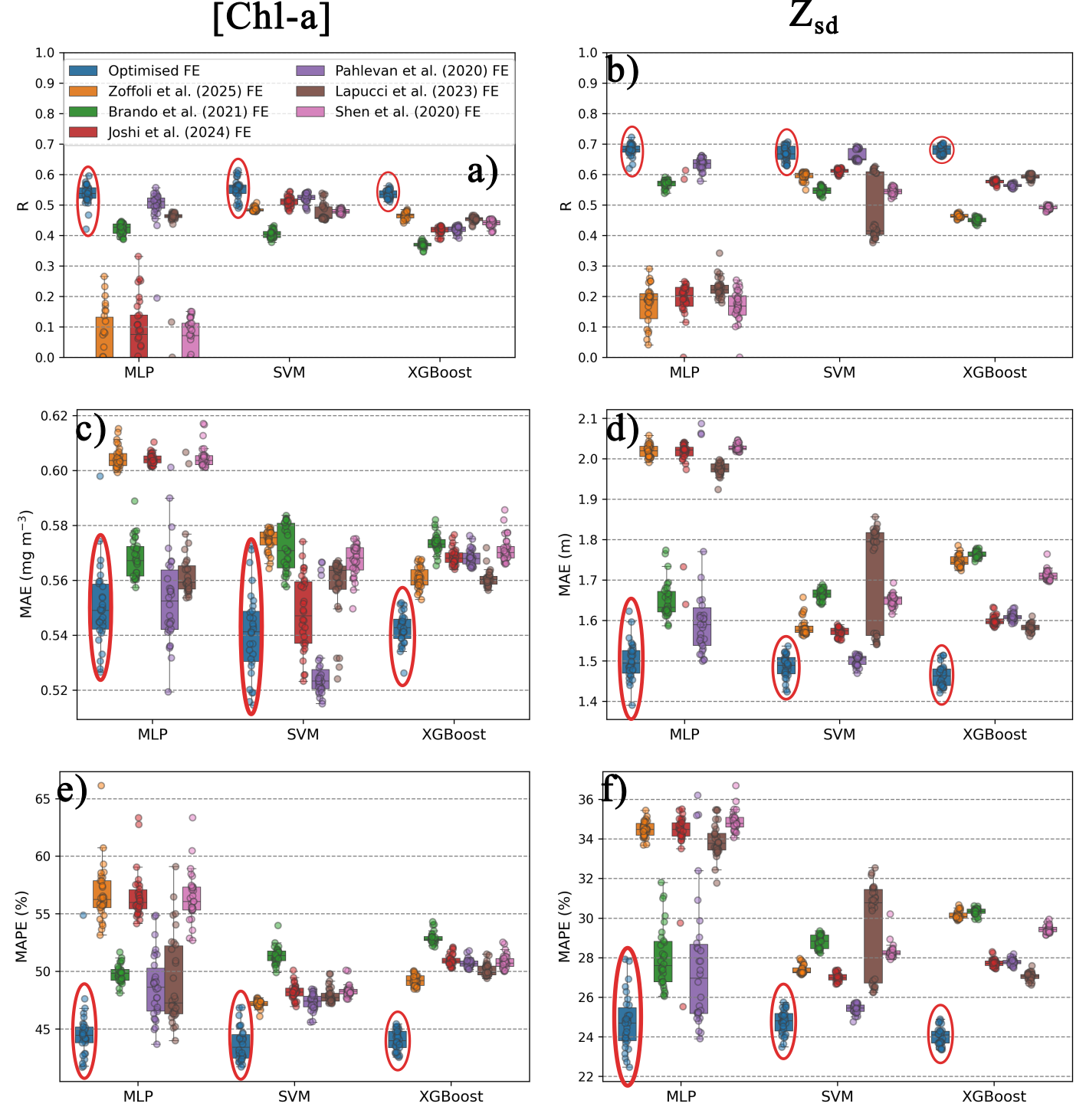}
\caption{Statistical results (R, MAE, and MAPE) for the FE experiments conducted. Subplots a, c, and e) are for [Chl-a] and b, d, and f) are for $Z_{\text{sd}}$. R is shown in subplots a and b), MAE is shown in subplots c and d), and MAPE is shown in subplots e and f). The box plots represent the distribution over 30 TPE optimisations for each FE experiment, shown in different colours- The optimised FE are highlighted with red ellipses}
\label{fig:fig4}
\end{figure}

Using the FE available in the literature causes a large variability in the model's accuracy across all ML models and target variables (Figure \ref{fig:fig4}).
The R, MAE, and MAPE variability between FE choices is higher in the MLP models but still present in the SVM and XGBoost for both [Chl-a] and Z\textsubscript{sd}.
Furthermore, the accuracy variability across TPE optimisation runs (n=30) is higher (wider box-plots) in MLP models than SVM and XGBoost, indicating that the MLP configurations may remain unoptimized after 500 trials in some instances.
Once the FE is optimised, the calibrated models reach the highest accuracy in both [Chl-a] and Z\textsubscript{sd}.
Remarkably, \cite{pahlevan_seamless_2020} FE reaches accuracy close to the optimised FE for SVM.

\subsection{Optimised FE choices}
\begin{figure}[!]
\centering
\includegraphics[width=0.85\textwidth]{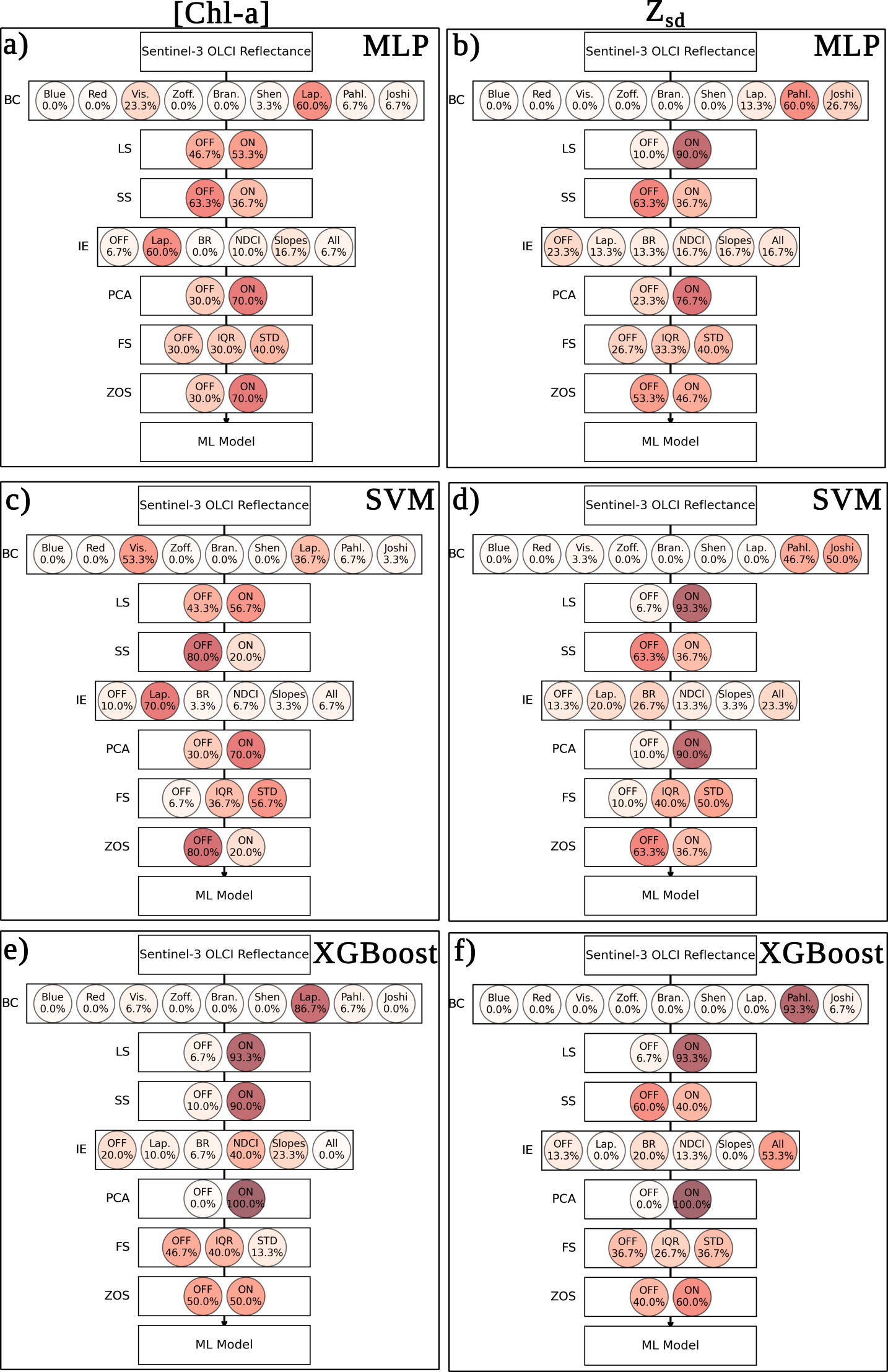}
\caption{Distribution of optimised FE choices. Results for [Chl-a] are shown in subplots a, c, and e) and for $Z_{\text{sd}}$ in b, d, and f). FE choices are shown for MLP in subplots a and b), for SVM in subplots c and d), and for XGBoost in subplots e and f). The percent values in each choice are the relative frequency of that choice, estimated as the best option across the 30 TPE optimisations. For example, 93.3\% of LS in ON for $Z_{\text{sd}}$ XGBoost models (in d, e, and f) corresponds to 28 of 30 runs}
\label{fig:fig5}
\end{figure}

The optimised FE exhibits irregular patterns across ML models and targets (Figure \ref{fig:fig5}).
In the [Chl-a], all ML models' band choices level converge to visible options (Vis. and Lap.) in more than 90\% of optimisation runs (n=30).
The Z\textsubscript{sd} models bands' choice level converges mostly to visible with NIR options (Pahl. and Joshi).
Log scaling level converges to ON for most of Z\textsubscript{sd} ML models and [Chl-a] XGBoost, while indifferent (close to 50\%) for [Chl-a] MLP and SVM.
The spectral shape level convergence exhibits no clear pattern, being predominantly ON and OFF or indifferent in several instances.
The convergence in the index extraction level seems indifferent for the Z\textsubscript{sd} ML models, except for XGBoost, where it converges to the All option in 53.3\% of runs.
For [Chl-a], the index extraction level for MLP and SVM converges to Lapp. option while XGBoost converges more often to the NDCI option in 40\% of the runs.
The PCA level converges to ON in most cases and up to 100\% XGBoost runs for [Chl-a] and Z\textsubscript{sd}.
Lastly, the feature scaling and zero-to-one scaling exhibit no clear pattern among the target and ML models.

\subsection{Improvement against standard OC retrieval models}
\begin{figure}[t]
\centering
\includegraphics[width=0.9\textwidth]{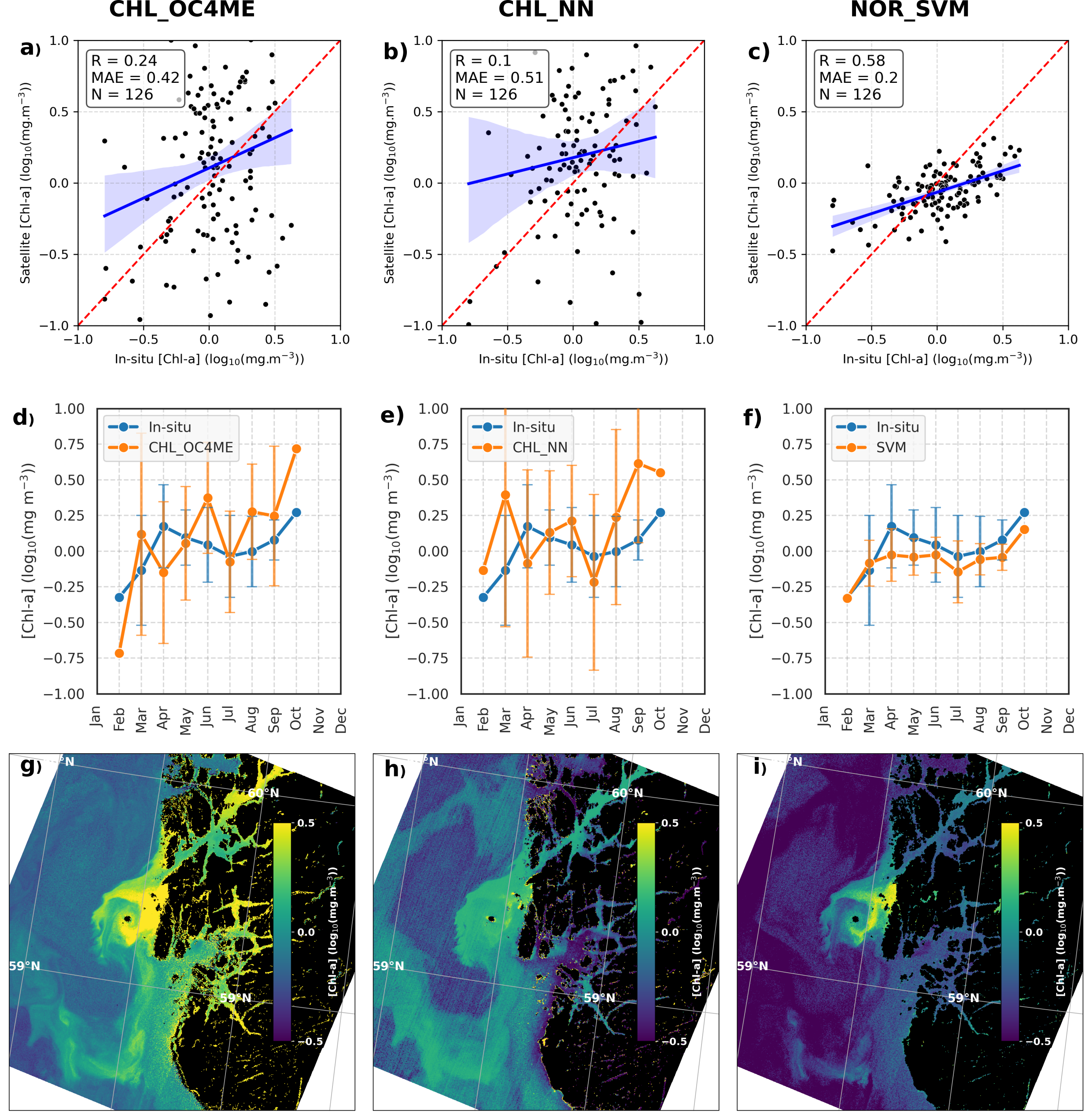}
\caption{Sentinel-3 OLCI [Chl-a] estimations. Subplots a–c) exhibit the matching of in-situ samples with the estimates by a) standard CHL\_OC4ME, b) CHL\_NN, and c) our NOR\_SVM algorithms.
Note that fewer samples are matched here because some points are masked as invalid data from the standard L2B OC product.
The matching observations aggregated in seasonal time series are shown for d) standard CHL\_OC4ME, e) CHL\_NN, and f) our NOR\_SVM algorithms, where dots are the monthly averages and bars represent standard deviations.
The retrieved maps (2023-06-15) of [Chl-a] distribution are shown for g) standard CHL\_OC4ME, h) CHL\_NN, and i) our NOR\_SVM. NOR\_SVM corresponds to the average of the 30 SVM models calibrated in this study}
\label{fig:fig6}
\end{figure}

Sentinel-3 OLCI standard retrieval models – CHL\_OC4ME and CHL\_NN – exhibit limited accuracy for estimating [Chl-a] in the Norwegian coastal waters (Figure \ref{fig:fig6}).
Considering both options, the OC4ME algorithm that is designed for Case-1 waters demonstrates slightly better results than the CHL\_NN that is designed for turbid waters.
Compared with the OC4ME, the average estimated by the 30 SVM models (referred to as NOR\_SVM) increases the R by 2.4 times and reduces the MAE by 52\%.
When aggregating the match-ups in a seasonal time series (Figure \ref{fig:fig6}d-f), in-situ observations exhibit a [Chl-a] peak in April and a later peak in October, which are the timing of the spring and autumn blooms of high latitudes \citep{silva_twenty-one_2021}.
The CHL\_OC4ME and CHL\_NN algorithms fail to reproduce such a seasonal pattern and show high-frequency variability along the months.
Our new NOR\_SVM, on the other hand, better demark the spring and autumn blooms similarly to in-situ observations, although with a slight underestimation.
Furthermore, NOR\_SVM demonstrate a high geographical variability with low [Chl-a] estimated in the open ocean and part of the inner fjords, moderate [Chl-a] in the inner fjords and in the cyan waters, and high [Chl-a] in the green waters (Figure \ref{fig:fig6}i).
CHL\_OC4ME estimates high [Chl-a] in most inner fjord areas where cyan blooms and dark waters are present.
CHL\_OC4ME has an overall high value that is probably related to the overestimation due to CDOM and particles (e.g., CaCO\textsubscript{3} from coccolithophore).
Both non-pigment constituents reduce the blue and green ratios that lead to Case-1 water algorithms to overestimate [Chl-a].
CHL\_NN, on the contrary, avoid the overestimation issue as it is designed for turbid and coastal waters. However, the CHL\_NN fails to estimate areas of high [Chl-a] in the green waters.

\begin{figure}[t]
\centering
\includegraphics[width=0.9\textwidth]{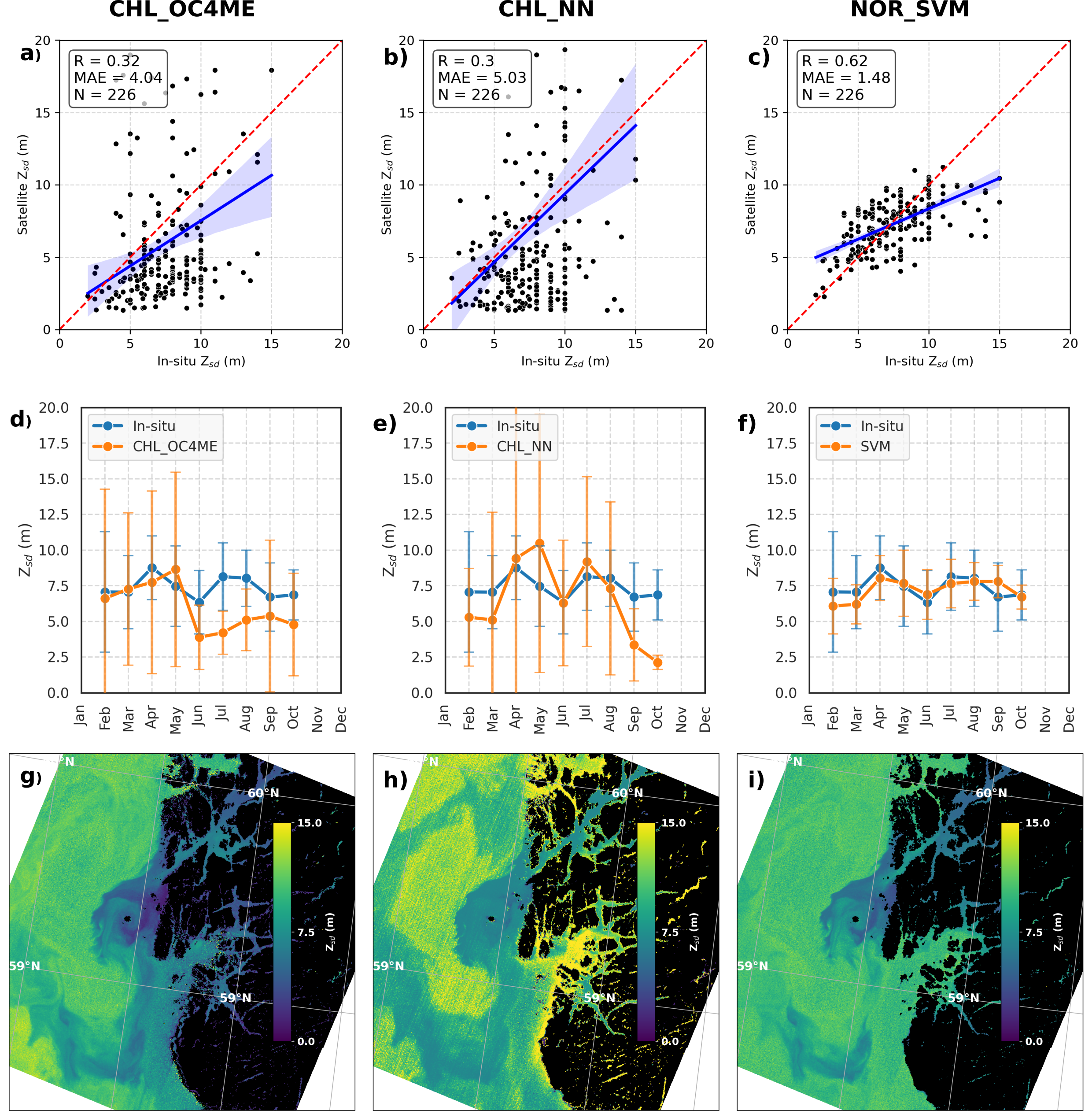}
\caption{Sentinel-3 OLCI $Z_{\text{sd}}$ estimations. Subplots a–c) exhibit the matching of in-situ samples with the estimates by a) standard CHL\_OC4ME, b) CHL\_NN, and c) our NOR\_SVM algorithms. Note that fewer samples are matched here because some points are masked as invalid data from the standard L2B OC product. The matching observations aggregated in seasonal time series are shown for d) CHL\_OC4ME, e) CHL\_NN, and f) NOR\_SVM algorithms, where dots are the monthly averages and bars represent standard deviations. The retrieved maps (2023-06-15) of $Z_{\text{sd}}$ distribution are shown for g) standard CHL\_OC4ME, h) CHL\_NN, and i) our NOR\_SVM algorithms. NOR\_SVM corresponds to the average of the 30 SVM models calibrated in this study, and CHL\_OC4ME and CHL\_NN use Eq.~1 with their estimated [Chl-a] as input for retrieval of the Secchi disk depth}
\label{fig:fig7}
\end{figure}

The Z\textsubscript{sd} derived by the equation \ref{eq:eq1} using the standard CHL\_OC4ME and CHL\_NN also demonstrates limited accuracy along the Norwegian coast (Figure \ref{fig:fig7}).
Hence, the NOR\_SVM substantially increase R by 1.9 times and reduces MAE by 63\% when compared to the CHL\_OC4ME.
The seasonal average in-situ Z\textsubscript{sd} exhibits peaks in April and in July to August that probably result from variations in incident light and light attenuation.
Next to the winter months (e.g., February and October), the incident light on the ocean surface is low as the sun zenith angle is high, which might result in the low Z\textsubscript{sd}.
During the summer, CDOM increases due to river runoff \citep{frigstad_influence_2020} and might explain the reduction observed in May and July.
CHL\_OC4ME fails to estimate this pattern and CHL\_NN slightly describes the same in-situ pattern, but with much larger seasonal variations.
On the other hand, the NOR\_SVM provides a closer relationship with the pattern observed in-situ.
Since Z\textsubscript{sd} using CHL\_OC4ME and CHL\_NN are derived from the [Chl-a] estimations, the spatial distributions follow a similar pattern.
The CHL\_OC4ME estimates shallow Z\textsubscript{sd} in the inner fjords associated with dark, cyan, and green waters.
The standard CHL\_NN estimate deeper Z\textsubscript{sd} in most regions, including inner fjords, as a result of the [Chl-a] underestimation.
The NOR\_SVM estimates shallow Z\textsubscript{sd} mostly in the green and cyan waters, while dark and blue waters exhibit deeper Z\textsubscript{sd}.

\section{Discussion}
Regional bio-optical algorithms often perform better than the ones designed to open oceans (Case-1) or to other coastal areas with contrasting optical variability.
\cite{tessin_testing_2024} has previously demonstrated this limitation in the Norwegian coastal waters for Sentinel-3 OLCI products and we corroborate the same result for our match-up data.
The weak performance of standard Sentinel-3 OLCI algorithms might be due to high optical variability and the prevalence of low reflectance waters dominated by CDOM.
The calibration of regional algorithms shows up as necessary and the ML usefulness for solving this optimisation task will continue gaining popularity as the records of public field monitoring data and satellite observations continue growing.
The increasing use of ML will benefit from know-how practices that improve the optimisation of ML models.
One example is including the day of the year with $R_{rs}$ that may help the ML model to account for seasonal variations in bio-optical relationships \citep{zoffoli_ciao_2025}.
Another example is employing an ensemble of ML models with varying wavelengths that can account for the temporal and spatial variation of atmospheric correction uncertainties \citep{brando_phytoplankton_2021}.
To the extent of our knowledge, feature engineering is so far explored ad hoc and left as an initial step in data pre-processing.
However, our experiment demonstrates that the FE plays a large role in the accuracy of ML models.
When adapting FE from previous studies to our experiments in the Norwegian coastal waters, we demonstrate that accuracy variability in terms of R, RMSE, and MAPE is higher among the FE configurations than across three different ML architectures.
Since FE can span a variety of options, we propose a framework that can optimise FE with similar rigour to that used for calibrating ML hyperparameters.
Our study demonstrates that employing TPE across different ML models and target variables for optimising FE yields more accurate models than using FE available in the literature.

The main challenge --- similar to hyperparameters --- is that no common optimal FE is found for all cases.
Our results suggest that the optimal FE options depend on the ML model and target variable.
Despite some FE optimisations exhibiting some common convergences, such as log-scaling reflectance to Z\textsubscript{sd} models and maximising the number of visible bands selected for [Chl-a] models, this outcome may depend on the regional optical conditions.
The surface reflectance broadly varies along the Norwegian coast as it is composed of waters with high absorption caused by CDOM and high scattering caused by coccolithophore blooms, both contributing to the reduction of water transparency and Z\textsubscript{sd}.
Log scaling might help allocate the high reflectance variability to a normal distribution and support the training of ML models.
For [Chl-a], on the other hand, log scaling reflectance seems less important as it is only found relevant for XGBoost.
Furthermore, high NIR reflectance commonly observed in eutrophicate environments \citep{cairo_hybrid_2019} is not present in our match-ups since the measured [Chl-a] are relatively low.
Hence, the NIR bands are probably a source of noise rather than something useful for estimating [Chl-a] using ML models in our region.
This might explain the band choice optimisations converging mostly to visible bands without the NIR being included.
In an eutrophicate environment, optimising FE might include the NIR bands.

A common criticism made against ML models is whether physical meaning exists from the optical input (e.g., reflectance) to the target variable (e.g., [Chl-a] and Z\textsubscript{sd}).
While semi-empirical and semi-analytical models exhibit a more transparent relationship \citep{lee_deriving_2002, morel_examining_2007}, ML models transform the data in such a way that might limit the physical understanding.
Optimising FE can, to some extent, restrict this understanding, as reflectance might be transformed several times prior to feeding the ML model.
For example, the relationship between reflectance and [Chl-a], where the original reflectance is sequentially log-scaled, has its principal components extracted, scaled by their variance, scaled from zero to one, and transformed by the given ML model architecture.
Despite this drawback, employing such transformations yields substantially improved estimations in coastal waters, and it should not be easily discarded.
Besides, we demonstrate that despite the sequence of transformations, we can still correct common issues in coastal waters and represent well the expected natural variability. 

The main concern in our region is the [Chl-a] overestimation caused by high CDOM that can lead to geographical and seasonal biases.
For example, the Case-1 waters CHL\_OC4ME algorithm estimates excessively high [Chl-a] in many areas where the NOR\_SVM estimate low values.
The CHL\_OC4ME led to several false positives of high biomass algae blooms, which can potentially mislead the detection of harmful ones that occur in Norway \citep{pettersson_monitoring_2013, john_spatial_2022}.
Although the CHL\_NN is advocated as an alternative for coastal waters, its accuracy performs worse than CHL\_OC4ME along the Norwegian coast and fjords, where our match-up data is located.
Conversely, our calibrated models correct the overestimation issue and provide a more realistic geographical and seasonal variability of [Chl-a] --- as well as Z\textsubscript{sd} --- in the waters they are trained for.
The NOR\_SVM slightly underestimate the high and overestimates the low [Chl-a] observed in situ, but this is probably caused by the smoothing of the grid cell area in satellite observations.
The same issue is also observed in the open ocean of the Norwegian Sea when using the Ocean Colour Climate Change Initiative (OC-CCI) data \citep{silva_twenty-one_2021}.
Despite the smoothing of [Chl-a], the NOR\_SVM can distinguish well the bloom areas from oligotrophic regions along the coast.

\section{Conclusion}
We introduce a framework for optimising FE and improving the accuracy of ML ocean colour coastal products.
The framework is demonstrated in the Norwegian coastal waters, which depicts a wide panel of challenges met: Atmospheric pollution, complex coastline, and high optical variability and complexity.
We show that FE can be more important than the pure ML model choice.
The TPE search algorithm optimises the FE configuration among many options drawn in an FE space by using the same rigour as for hyperparameter tuning.
The FE space works as a pipeline where the input data (reflectance) flows through feature selection and data transformations applied in seven sequenced levels.
Although our FE space is drawn by adapting the current literature of ML models applied to Sentinel-3 OLCI, we expect that future applications are not confined to those.
Expanding the FE space may be used to improve our framework in the future, in particular, expanding the domain-specific FE by harvesting the extensive ocean colour literature.
For example, the FE space could integrate atmospheric correction, glint removal, adjacency correction, and semi-analytical models.
Those potential new FE levels could explore different techniques and the parameters belonging to them.
Considering the continuing growth of observational databases and following the use of ML models, optimising FE can be key for the calibration of accurate regional retrieval models in coastal areas.
Hence, contributing to making the ocean colour data more reliable for environmental ecosystem assessment, also in optically complex coastal and fjord waters.

\begin{Backmatter}
\paragraph{Acknowledgments}
We appreciate the availability of in situ data from the national Økosystemovervåking i kystvann (Økokyst) from the Norwegian Environment Agency and the Sentinel-3 OLCI data from the Copernicus Data Space.

\paragraph{Funding Statement}
ES and SC were funded by the Bjerknes Centre for Climate Research (BCCR) – Centre of Climate Dynamics (SKD) Strategic Project NorHAB-ML and by the institute research fellowship (INSTSTIP) funded by the basic institutional funding through the Norwegian Research Council (\#318085).
LP was funded by the Research Council of Norway through the Nansen Center basic funds grant no. 342624.
FC acknowledges the NFR Climate Futures (309562).
This work has also received a storage Grant (NORSTORE, NS9039K)

\paragraph{Competing Interests}
None

\paragraph{Data Availability Statement}
Chlrophyll-a and Secchi disk depth data are publicly available on the Vannmiljø web portal (https://vannmiljo.miljodirektoratet.no/). Sentinel-3 OLCI observations are publicly available on Copernicus Data Space Ecosystem (https://dataspace.copernicus.eu/). Codes for reproducing the study will be made available once accepted for publication.

\paragraph{Ethical Standards}
The research meets all ethical guidelines, including adherence to the legal requirements of the study country.

\paragraph{Author Contributions}
Conceptualisation: E.S.,
Data Curation: E.S.
Formal analysis: E.S., S.C., J.B.
Funding Acquisition: J.B., F.C., L.P.
Investigation: E.S., J.B., S.C., F.C., L.P.
Methodology: E.S., J.B.
Validation: E.S., J.B., S.C., F.C., L.P.
Visualisation: E.S.
Writing - Original Draft: E.S.
Writing - Review and Editing: E.S., J.B., S.C., F.C., L.P.

\paragraph{Supplementary Material}
None.

\bibliography{references}
\bibliographystyle{apalike}

\end{Backmatter}
\end{document}